# Isomer triplets in odd-odd transitional rare earth nuclei: unique features, orbital systematics and characterization

N. Susshma, S. Deepa, K. Vijay Sai and R. Gowrishankar

*Department of Physics, Sri Sathya Sai Institute of Higher Learning, Prasanthi Nilayam-515134, INDIA*

**Abstract**

The existence of low-lying long-lived isomers, predominantly in odd-odd nuclei of the light rare-earth mass region, is investigated through an extensive survey of available nuclear data. The characteristics of these isomeric states and their systematics has revealed intriguing and unusual properties, including the identification of 'isomer triplets' - a phenomenon specific to odd-odd deformed nuclei close to the transition region. This exclusive feature was observed in the following odd-odd nuclei– $^{152}$Pm, $^{152}$Eu, $^{154}$Tb, $^{156}$Tb, $^{156}$Ho, $^{158}$Ho, $^{160}$Ho, $^{162}$Lu and $^{166}$Lu. We present a detailed overview of these isomer triplets by exploring the systematics of single-quasiparticle (1qp) proton and neutron orbitals near the Fermi surface relevant in this mass region, to elucidate the factors responsible for their formation. The low-lying level structures of $^{156}$Tb and $^{154}$Tb, were constructed using the well-tested Two Quasiparticle Rotor Model to resolve the ambiguities in the spin, parity, energy and orbital configuration of these isomeric states. These results were extended to study the systematics of low-lying isomer triplets in the other five light rare-earth nuclei of interest. Our review and analysis of 1qp proton and neutron orbital systematics in the neighbouring odd-mass isotopes and isotones highlights the crucial role of high-spin intruder neutron orbital in the formation of these isomer triplets.

## I. Introduction

The rotational deformed mass region (150<A<190) is known to consist of nuclei with intriguing, exotic properties which are manifestations of their deformed shapes. Experimental and theoretical studies are constantly attempting to expand the understanding of these non-spherical shapes, their complex excitations and the resulting level structures. Many nuclei in this region are observed to exhibit isomeric states, shape coexistence, shape transition, triaxiality, octupole correlation etc. The unique properties of deformed rare-earth and actinide nuclei were first demarcated by the collective shell model and its level structures are best defined by the Nilsson model [1,2]. The degenerate energy levels defined by the angular momentum projection quantum number Ω [Nn$_3$ΛΣ] relative to deformations realistically describe the level schemes of deformed nuclei. In the rare-earth nuclei, the degenerate Nilsson levels of $h_{11/2}$, $f_{7/2}$, $h_{9/2}$ and $i_{13/2}$ occur close to the Fermi surface. The multi-quasiparticle excitations in even-even nuclei and the outermost orbital occupied by the unpaired proton/neutron in odd-mass nuclei are defined by the Nilsson energy levels. In case of odd-odd nuclei, the two-quasiparticle (2qp) states are formed by the coupling of unpaired proton and neutron levels.

The relatively simpler structure formation of even-even and odd-mass nuclei enabled many experimental explorations of their exotic properties. Islands of shape coexistence were found at N=90 and N=60 rare-earth even-even isotones [3]. Triaxiality has been experimentally observed through wobbling bands in the odd-mass rare-earth nuclei $^{151}$Eu, $^{163}$Lu, $^{167}$Ta, and $^{183}$Au [4–7], to name a few. Isomeric states with half-lives ranging from few nanoseconds to thousands of years are found in a large number of rare-earth nuclei [8,9]. Though isomers and their half-lives were one of the earliest experimentally studied nuclear phenomena, their discovery and especially understanding the reason behind their formation have greatly contributed in the development of nuclear models and thereby, the overall understanding of nuclear structure. Observation of the first K-isomers in $^{178}$Hf and $^{180}$Hf led to the earliest proposition of rotational deformations in even-even nuclei [1,10,11].

Isomers are classified into different types based on the phenomena causing their formation broadly as spin isomers, shape isomers, seniority isomers and K-isomers [8]. Further classifications of shape isomers into fission isomers and spin isomers into seniority isomers were delineated owing to advancement in experimental and theoretical model-based studies of nuclear structure [9,12]. There have been several extensive survey reports/compilations of isomers across mass regions, wherein their categories and the underlying causes for their occurrence have been outlined [12–15]. The periodically updated 2023 Atlas of Isomers [13] and Maheshwari et. al. [15] reported maximum number of long-lived isomers in odd-odd nuclei across the nuclear landscape. However, the reports majorly focus on spin and shape isomers in odd-mass and even-even nuclei near shell closures, K-isomers in A=180 region, and fission and extreme low energy isomers in superheavy nuclei. In an earlier report in the medium-heavy ($Z \geq 65$ and $N \geq 91$) and heavy deformed nuclei, Sood & Sheline [16] compiled the existing isomers in odd-mass and odd-odd nuclei and touched upon the role played by single particle proton and neutron states in their formation. A more recent report by Orford et. al. [17] was an experimental determination of isomers in the $95 \leq N \leq 101$ odd-odd Pm, Eu and Tb isotopes. The work confirmed the existence of low energy 'spin-trap' isomers in the light rare-earth nuclei around N~98 caused by the large spin difference in 2qp Gallagher Moszkowski (GM) doublets in the low-level scheme.

Earlier reported surveys of odd-odd nuclei in the medium-heavy (A=144-194) mass region [18,19] show abundant existence of such low-lying long-lived isomers (LLIs). While these works have analysed and reported the isomers and their formation in many sets of nuclei, there is lack of detailed study on trends of low-lying isomers close to the transitional region. While the current data sheets [20] report the dense existence of low-energy isomers, a thorough survey of long-lived isomers in doubly-odd deformed nuclei of this region has revealed a few other unique and interesting features. Our present study exclusively reports the observation and systematics of 'isomer triplets', which are a set of three low-lying (E< 500 keV) long-lived ($t_{1/2} \geq 1s$) energy levels, comprising of the ground state and two LLIs with somewhat comparable half-lives. Occurrence of such a trend of isomeric clusters in the light and medium mass (150<A<170) rare-earth nuclei is being reported explicitly for the first time. A compilation of nine such rare-earth nuclei with isomer triplets and the available data on those individual states is presented in Table 1. It can be seen that the energies, spins and parities of many of these isomers remain uncharacterised. There are also possible occurrences of such triplets in $^{172}$Ir and $^{176}$Au, which we have not listed since the experimental data lacks information on the ground states of these nuclei. On the high mass end of the deformed region, $^{190}$Ir is reported to have a low-lying isomer triplet [20]. $^{186}$Ta is expected to have an isomer triplet [21] though they have so far not been experimentally confirmed.

***Table 1:*** *A compilation of isomer triplets known to exist from reported experimental studies [20] in the transitional light rare-earth region. The half-life, excitation energy, spin, parity and percentage of decay, experimentally known so far, are also listed alongside.*

| $^A$X | $t_{1/2}$ | $E_x$ (keV) | $J^\pi$ | %β | %IT |
|---|---|---|---|---|---|
| $^{152}$Pm | 4m | 0 | $1^+$ | 100 | |
| | 8m | $1.5 \times 10^2$ | $4^-$ | 100 | -- |
| | 14m | 150+x | (8) | ≤100 | ≥0 |
| $^{152}$Eu | 13y | 0 | $3^-$ | 100 | |
| | 9h | 46 | $0^-$ | 100 | |
| | 96m | 148 | $8^-$ | -- | 100 |
| $^{154}$Tb | 21h | 0 | 0 | | |
| | 9h | 0+x | $3^-$ | 78 | 22 |
| | 23h | 0+y | $7^-$ | 98 | 2 |
| $^{156}$Tb | 5d | 0 | $3^-$ | 100 | |
| | 24h | 50+x | $(7^-)$ | -- | 100 |
| | 5h | 88 | $(0^+)$ | >0 | <100 |
| $^{156}$Ho | 56m | 0 | $4^-$ | 100 | |
| | 9s | 52 | $1^-$ | | 100 |
| | 8m | 52+x | $9^+$ | 75 | 25 |
| $^{158}$Ho | 11m | 0 | $5^+$ | 100 | |
| | 28m | 67 | $2^-$ | <19 | >81 |
| | 21m | 180(CA) | (9$^+$) | ≥93 | ≤ 7 |
| | 26m | 0 | 5$^+$ | 100 | |
| $^{160}$Ho | 5h | 60 | 2$^-$ | 24 | 76 |
| | 3s | 170+x | (9$^+$) | -- | 100 |
| | 1.4m | 0 | 1$^-$ | 100 | |
| $^{162}$Lu | 1.5m | X | (4$^-$) | ≤100 | - |
| | 1.9m | Y | - | ≤100 | - |
| | 2.65m | 0 | 6$^-$ | 100 | |
| $^{166}$Lu | 1.41m | 34 | 3$^{(-)}$ | 58(5) | 42(5) |
| | 2.12m | 43 | 0$^-$ | >80 | < 20 |

The distinctive features of the data compiled in Table 1 are as summarised below:

a) These triplets are observed only in lighter mass rare earths overlapping the transitional region. Presently identified triplets include only nuclei with N=89/91/93.
b) Almost in all cases, the three isomers in each triplet have comparable half-life.
c) In case of neighbouring isomers with ΔI=3, the E3/M3 transition is greatly hindered, or even absent.

In the absence of unambiguous experimental data, theoretical model-based studies provide crucial insights, acting as location guides for understanding the properties of the nuclei of interest. In our exploration of low-lying isomer triplets in the light rare-earth odd-odd nuclei, we have used the well-tested empirical Two Quasiparticle Rotor Model (TQRM) calculations to obtain their low energy level scheme. Section II of the present study discusses briefly the framework of this theoretical model and the formulae used to calculate spin-parities, energies and orbital configurations of the low lying 2qp states. The Tb isotopes listed in Table 1 have been studied in detail using the TQRM in Section III. The model calculated energies and J$^\pi$ assignments of the physically admissible 2qp band heads in $^{154,156}$Tb are compared with the existing experimental data thereby providing explanations for ambiguities in the latter. The trend of isomer triplet states in odd-odd Tb isotopes in the mass region 150 ≤ A ≤ 160 are compared and analysed to understand the cause of their formation. In section IV, the results of the Tb isotopes are extended to the isomer triplets in the other light rare-earth nuclei listed in Table 1. This process has led to the identification of the systematics of one quasiparticle (1qp) proton and neutron orbitals near the Fermi surface and their significant role in the formation of these LLIs.

**II. Outline of the TQRM formulation**

The empirical TQRM is a three-step process to deduce the level energies and 2qp configurations of the low-lying spectra of odd-odd nuclei [18,19,22]. The first step is to identify the 1qp proton and neutron configuration space within a specified energy range by examining the systematics of experimental excitation spectra of the nearest neighbour odd-A isotope and isotone. The second step is to obtain the physically admissible 2qp states arising from the coupling of the 1qp proton and neutron orbitals. For the odd-odd nucleus, the coupling of unpaired proton ($\Omega_p$) and neutron ($\Omega_n$) gives rise to two states, with spins K$^+$= $\Omega_p$+$\Omega_n$ and K$^-$= $\Omega_p$− $\Omega_n$, known as the Gallagher-Moszcowski (GM) doublet. The GM rule [23] places the spins anti-parallel triplet state (K$_T$) at a relatively lower energy than the spins parallel singlet state (K$_S$). The final step is to calculate the 2qp bandhead energies using the following expression [18,22]:

$$E(\Omega_p, \Omega_n) = E_0 + E_p(\Omega_p) + E_n(\Omega_n) + E_{rot} + <V_{pn}> \quad (1)$$

Here, E$_p$ and E$_n$ are the experimentally observed energies of the 1qp proton and neutron Nilsson levels taken from the neighbouring isotope and isotone respectively. The rotational energy E$_{rot}$ is calculated from the lower spin number ($\Omega_<$) of the 1qp proton or neutron levels forming the doublet as

$$E_{rot} \approx \frac{\hbar^2}{2I}[K^\pm - (\Omega_p + \Omega_n)] = \frac{\hbar^2}{2I}(2\Omega_<)\,\delta_{K,K^-} \quad (2)$$

The term $<V_{pn}>$ is the contribution of the residual n-p interaction including the GM splitting energy $E_{GM}$ and the Newby shift energy $E_N$ which is present only in case of K=0 bands.

$$<V_{pn}> = -(\frac{1}{2} - \delta_{\Sigma,0}) E_{GM} + (-)^I E_N \delta_{K,0} \qquad (3)$$

It can be seen that the $E_{GM}$ term is negative for the spins-antiparallel $K_T$ and positive for the spins-parallel $K_S$ state. In the semi-empirical approach, we use the experimentally determined values of $E_{GM}$ and $E_N$ for the specific 2qp configuration from neighbouring odd-odd nuclei, considering that the energy parameters are configuration specific and not nucleus specific. Over the last several years, this simplified TQRM formulation has been effectively employed to describe and to predict the location and character of 2qp bands and occurrence of long-lived low-lying isomers in various odd-odd deformed nuclei in both the rare earth and actinide regions [21,24–30]. Detailed discussions of the TQRM formulation can be found in any of the above references.

### III. Isomers of $^{154,156}$Tb

We begin with the survey of experimental single particle Nilsson orbitals for protons and neutrons in the neighbouring isotopes and isotones relevant to $^{154,156}$Tb. This orbital systematics, in the mass region A=150 - 160, are plotted in Fig. 1. We observe some intriguing trends as elucidated here. The ν3/2[521↑] orbital predominantly characterises the g.s. of N=91 isotones $^{155}$Gd, $^{157}$Dy and $^{159}$Er, while otherwise found at higher energy in the lower mass isotones $^{151}$Nd and $^{153}$Sm. In Gd isotopes, this n-orbital consistently remains the g.s. in the 152<A<160 mass region. Orbitals of the next highest energies lie very close to each other throughout the A=153-159, as seen in figure 1. As noted by the evaluators, in $^{155}$Gd, the $J^\pi$=3/2$^+$ level at 105 keV is an admixture of ν3/2$^+$[651↑] majorly with significant components of the $i_{13/2}$ related Nilsson state as well as ν3/2$^+$[402↓] orbital caused by strong ΔN=2 mixing. Additionally, experimental data confirm that the 5/2$^+$ rotational level of ν3/2[651] band lies lower in energy ($E_x$=87 keV) than its 3/2$^+$ ($E_x$=105 keV) bandhead in $^{155}$Gd. We may also note the dipping of high spin ν11/2[505↑] neutron orbital to lower energies only in $^{153,155}$Gd isotopes in contrast to the neighbouring Gd isotopes. As for the p-orbitals of interest, we observe a crossover between of 5/2$^+$[402↑] and 3/2$^+$[411↑] between $^{153}$Tb and $^{155}$Tb. The former is the g.s in $^{153}$Tb following which there is a sharp rise in its energy with increase in mass number. Whereas, the latter falls sharply from 150 keV and remains as g.s. for A=155,157 and 159 Tb isotopes. Based on this systematics, we have mapped the relevant 1qp p- and n- orbitals to construct the 2qp states in $^{154,156}$Tb.

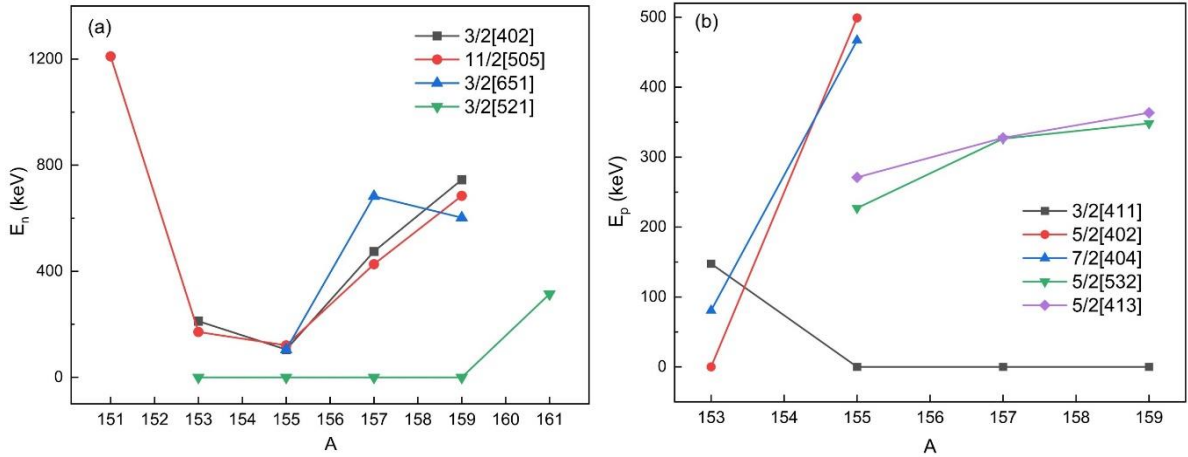

*Figure 1:* (a) The experimental neutron orbital systematics of relevant 1qp orbitals in the A=150-160 mass region as taken from odd-mass Gd isotopes which are isotonic neighbours of the odd-odd Tb isotopes. The 3/2[402] orbital undergoes strong ΔN=2 non-coriolis mixing with 3/2[651] and has some components of $i_{13/2}$ Nilsson states at A=155. (b) The proton orbital systematics of low energy 1qp orbitals in the A= 150 to 160 mass region taken from the odd-mass Tb isotopes. There is a sharp increase/decrease of the proton orbital energies around A= 154.

### A. $^{156}$Tb

The physically admissible 2qp bandheads ($K_T$ & $K_S$) in $^{156}$Tb arising from the coupling of 1qp proton and neutron orbitals as taken from $^{155}$Tb and $^{155}$Gd respectively are listed in Table 2 up to an energy of $E_p + E_n$ = 550 keV. As evident therein, even in this low-energy domain, we expect 30 2qp bands while the presently available experimental data identifies only 6 rotational band structures. This situation is in sharp contrast to that in the isotonic neighbour $^{154}$Eu wherein 32 bands have been clearly identified. Our TQRM calculated bandhead energies are plotted in Fig. 2.

***Table 2:*** *The physically admissible 2qp band heads in $^{156}$Tb as determined from the single particle proton states of $^{155}$Tb and neutron states of $^{155}$Gd up to $E_x(=E_p+E_n) \sim 550$ keV*

| $p_j \rightarrow$ <br> $n_i \downarrow$ | $p_0$= 0 keV <br> $3/2^+$ [411↑] | $p_1$=227 keV <br> $5/2^-$[532↑] | $p_2$=271 keV <br> $5/2^+$ [413↓] |
|---|---|---|---|
| $n_0$= 0 kev <br> $3/2^-$ [521↑] | $3^-$   $0^-$ <br> (0) | $4^+$   $1^+$ <br> (227) | $1^-$   $4^-$ <br> (271) |
| $n_1$= 105 kev <br> $3/2^+$ [651↑] | $3^+$   $0^+$ <br> (105) | $4^-$   $1^-$ <br> (332) | $1^+$   $4^+$ <br> (376) |
| $n_2$=121 kev <br> $11/2^-$ [505↑] | $7^-$   $4^-$ <br> (121) | $8^+$   $3^+$ <br> (348) | $3^-$   $8^-$ <br> (392) |
| $n_3$=267 keV <br> $5/2^+$ [642↑] | $4^+$   $1^+$ <br> (267) | $5^-$   $0^-$ <br> (494) | $0^+$   $5^+$ <br> (538) |
| $n_4$=269 keV <br> $3/2^+$ [402↓] | $0^+$   $3^+$ <br> (269) | $1^-$   $4^-$ <br> (496) | $4^+$   $1^+$ <br> (540) |

1. ***The ground state GM doublet***

The $^{156}$Tb g.s. was assigned a $J^\pi$=3$^-$ from atomic-beam magnetic resonance studies [31]. Our analysis yields $J^\pi$=3$^-$ as the spins-parallel triplet state ($K_T$) of the g.s. GM pair ($p_0$:3/2$^+$[411↑] ⊗ $n_0$: 3/2$^-$[521↑]), in total agreement with the experimental data. While there are no reported experimental observations of its 2qp singlet partner $J^\pi$=0$^-$, our model calculations place this level at $E_x \approx$120 keV. Further, the $I^\pi$ = 4$^-$ member of the $K^\pi$=3$^-$ g.s. band is determined to be at energy $E_x \approx$100 keV, in close agreement with that proposed in the existing data sheet based on "comparison of the measured ($^3$He, d) and (α, t) cross sections with those predicted for the members of the band having the proposed configuration together with the expected energy spacing" [20].

2. ***The 5.3 h low spin isomer***

The $t_{1/2}$=5.3 h isomer, which is the second among the three low-lying isomers in $^{156}$Tb is experimentally found at $E_x$=88.4 keV based on an E3 transition from this level to ground state [32,33]. As discussed by Toriyama et. al. [34], and noted by the ENSDF evaluators [20], the spin of this level is expected to be less than the g.s. J=3 and hence this isomer is tentatively assigned a $J^\pi$= (0$^+$). The parallel state coupling of ($p_0$ ⊗ $n_4$) results in $K^\pi$=0$^+$, whose energy is expected to be $E_x \sim$90 keV due to the ΔN=2 mixing of the ν3/2$^+$[651] and ν3/2$^+$[402] orbitals as discussed in a preliminary report by Sood et. al. [35] Given the above, we confirm and assign the spin and 2qp configuration of the low spin isomer to be

**5.3 h $^{156}$Tb$^{m1}$: $J^\pi$=0$^+${π3/2$^+$[411] ⊗ ν3/2$^+$[402]}; $E_x \approx$ 90 keV**

3. ***The 24.4 h high spin isomer***

One of the significant open questions in the available data on $^{156}$Tb is the placement of the high-spin long lived ($t_{1/2}$=24.4 h) isomer and its decay route. Toriyama et. al. [34] first reported the existence of this isomer from the half-life measurements of the $E_\gamma$= 49.6 keV transition to the g.s. of $^{156}$Tb. Their analysis elucidates the decay route of this isomer through another transition of low-energy and high multipolarity above the 49.6 keV transition in cascade. They also suggested that the spin of this isomer to be J > 3. Bengstonn et. al. confirmed the $E_x$=49.6 keV level to be a short-lived isomer itself ($t_{1/2}$≈ 49 ns) apart from being fed by the 24.4 h isomer decay [36]. The current data sheets [20] list this LLI with a tentative assignment $J^\pi$ = (7$^-$) based on comparison with the g.s. and reported isomeric states in $^{154}$Tb. As seen in Table 2, the physically admissible high spin 2qp state at low energies results from the spins-parallel coupling of $p_0 \otimes n_2$ at $E_x \sim$ 130 keV. Other high spin 2qp states are expected at energies $E_x$> 250 keV. The above arguments lead us to categorically assign $J^\pi$ and configuration to this isomer along with its placement as:

**24.4 h $^{156}$Tb $^{m2}$:** 7$^-$ {$\pi$3/2$^+$[411] $\otimes$ $\nu$11/2$^-$[505]}, $E_x$≈130 keV

Our evaluation admits a possible decay to the 49.6 keV $J^\pi$=4$^+$ level by an E3 transition ($E_\gamma \approx$ 80 keV). It may be noted that this transition may have same energy as that of the reported $E_\gamma$=88.4 keV ($J^\pi$ =0$^+$ → 3$^-$g.s.) E3 transition, suggesting an unresolved doublet thereof.

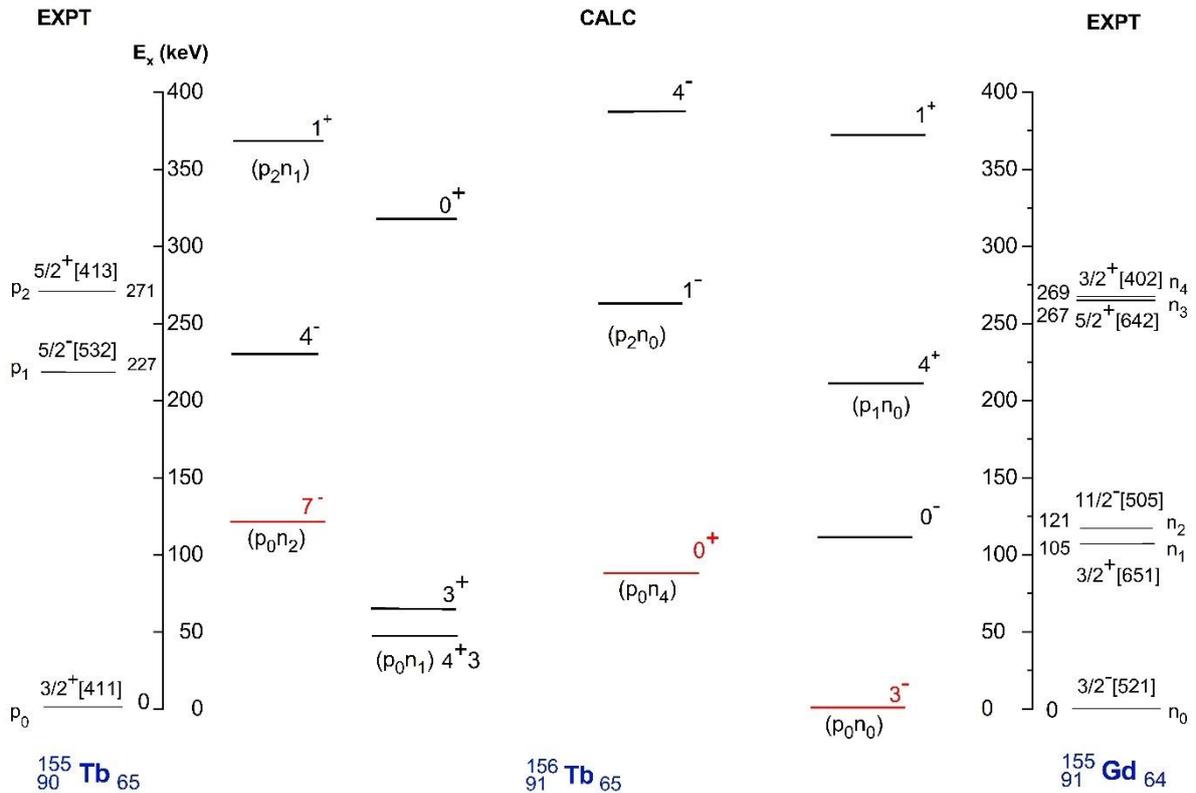

*Figure 2*: The 2qp band heads of $^{156}$Tb calculated using the empirical TQRM calculations up to orbital energy $E_x$=400 keV. The experimental energies and spins of 1qp orbitals were taken from $^{155}$Tb (proton levels) and $^{155}$Gd (neutron levels). The isomeric states (in red colour) have been observed and reported experimentally except for the $J^\pi$=7$^-$ whose spin and energy has been confirmed and proposed in the current study [20].

*4. Band heads of rotational bands*

Experimentally, apart from the ground state, two rotational bands with spin and energy assignments have been reported. These two bands were tentatively assigned configurations $\pi$3/2[411] $\otimes$ $\nu$i$_{13/2}$ and $\pi$h$_{11/2}\otimes$ $\nu$i$_{13/2}$ [36]. Among these two bands, the $\pi$3/2[411] $\otimes$ $\nu$i$_{13/2}$ was built on the short-lived 4$^+$ isomeric state at $E_x$=49 keV. The ENSDF evaluators proposed the likely configuration of the band to be $\pi$3/2$^+$[411] $\otimes$ $\nu$5/2$^+$[642]. However, from the 1qp n orbital systematics in the neighbouring isotone $^{155}$Gd, the 5/2$^+$[642] is observed at much higher energy

($E_x$=267 keV) and hence its coupling with π3/2[411] will not yield $K^\pi$=4$^+$ bandhead at energies as low as 50 keV. The $J^\pi$=4$^+$ state formed by the coupling of $p_0 \otimes n_1$ where the 5/2$^+$ rotational level of the 3/2$^+$[651] band is found to be lower than the 3/2$^+$ bandhead is found to be at $E_x$= 49 keV. We hence, propose an alternate configuration for the π3/2[411] ⊗ vi$_{13/2}$ band as $p_0$: π3/2$^+$[411] ⊗ $n_1$: v3/2$^+$[651]. A third rotational band was proposed from the $^{155}$Gd ($^3$He, d) (α, t) reaction studies [20] albeit the non-observation of any interband transitions. This band is listed starting from $I^\pi K$=1$^-$0 at $E_x$=100 keV with the assigned configuration π3/2[411] ⊗ v3/2[521]. It may be noted that its bandhead $K^\pi$=0$^-$ has not been experimentally identified. This bandhead could essentially be the singlet state of the g.s. GM pair, as discussed previously.

### B. $^{154}$Tb

Two low-lying isomers, along with the unstable long-lived g.s., and having $J^\pi$= 3$^-$ and 7$^-$ form the isomer triplet in $^{154}$Tb. These isomers were reported through multiple beta-decay studies [31,33,37–40]. The spin of the g.s. was assigned J=0 from a detailed decay activity studies by Lau & Hogan [41]. Nevertheless, the excitation energies of the other two isomers remain unknown over the past four decades [20]. From figure 1(b), we see a crossover of the π5/2[402] and π3/2[411] orbitals as we move from $^{153}$Tb to $^{155}$Tb. Also, the coupling of π5/2[402] with any of the low-lying n orbitals, as taken from $^{153}$Gd, will not yield spins of J=0 or 3 at low energies, which is in contradiction to the experimentally observed spins of J=0, 3 and 7 for the low-lying isomer triplet in $^{154}$Tb. Under these considerations, the π3/2[411↑] orbital is chosen as the lowest 1qp proton orbital ($p_0$) to construct the 2qp bandheads in $^{154}$Tb. In table 3, we list the physically admissible 2qp bandheads in $^{154}$Tb arising from the coupling of proton and neutron orbitals taken from the neighbouring odd mass isotope $^{155}$Tb and isotone $^{153}$Gd up to $E_p+E_n$=450 keV.

#### 1. *Ground state doublet and violation of GM rule*

Existing data sheet lists the g.s. of $^{154}$Tb as J=0 without any parity assignment. As seen in Table 3, the g.s. GM pair ($p_0 \otimes n_0$) is $K_T$ = 3$^-$ and $K_S$ = 0$^-$. According to Harmatz et. al. [42] where the J=3 and 0 isomeric states were first observed and reported, the Nilsson orbital assignment for these isomeric states was found to be 'not very definitive' due to the existence of $^{154}$Tb in the 'region of sharp change in nuclear deformation'. The parity of the J=0 state was proposed as either positive with a {π3/2[411↑] ⊗ v3/2[651↑]} configuration or negative with {π3/2[411↑] ⊗ v3/2[521↑]}. Some previous works like Sousa et. al. [37] assumed the parity to be positive, though no clear experimental evidence to substantiate it has been reported. Later Lau & Hogan [41] studied the decay activity of all three isomeric states, choosing γ-rays which specifically decay from each isomeric state. The study reported indirect evidence of an isomeric transition from the 3$^-$ state to the 0 state. This led to the accepted conclusion that J=0 should be the g.s. of $^{154}$Tb. The ambiguity on the parity however persists till date. Our detailed survey of the neutron orbital energies reveals that the v3/2[651↑] 1qp neutron orbital is at higher energy in comparison to the v3/2[521↑] throughout the A=150-160 mass region, the latter consistently remaining as the ground state as seen in Fig. 1(a). In view of this conclusive evidence, we categorically rule out positive parity for the 21.5 h g.s. and assign its configuration as $J^\pi$=0$^-$ {π3/2[411↑] ⊗ v3/2[521↑]}.

According to the GM rule [23], the spins parallel triplet state $K_T$=3$^-$ is expected to lie lower in energy than its spins antiparallel singlet state partner $K_S$=0$^-$ for the g.s GM doublet in $^{154}$Tb. However, in contradiction to this, experimental evidence suggests that the J=0 state is the g.s. in $^{154}$Tb, which is a clear violation of the GM rule. Jain et al. [18], in their detailed study of the nuclear structure in odd-odd rare-earth region, proposed possible GM rule violation in 12 rare-earth nuclei including $^{154}$Tb. The study predicts that, close to the transitional region, where the onset of rotational shape in nuclei begins (around A~150), the GM rule would no longer be valid. They suggest the quasiparticle-plus-phonon model as a better alternative in this case. The coupling of the external nucleon to the even-even core vibrations leads to appearance of vibrational admixtures in the low-lying states of odd-odd nuclei, which is generally not included in the TQRM analysis. The theoretical treatment of this additional component is described by the 2qp-plus-phonon model, wherein the intrinsic Hamiltonian includes the vibrational contribution of the even-even core [18]. Valantanin-Bogolyubov transformation is used to obtain the 2qp operators from the 1qp operators and Random Phase Approximation is used to define the intrinsic Hamiltonian. The p-n interaction potential ($V_{pn}$), GM splitting energy and Newby shift are obtained from defining the intrinsic wavefunction of an odd-odd state using the 2qp particle creation operator and thereby the intrinsic excitation energy of the state. Jain et. al. [18], using the 2qp-plus-phonon model calculated and reported the energy of the $J^\pi=3^-$ level to be ~12 keV in $^{154}$Tb. Our empirical evaluation of energy separation between the g.s. doublet using known energies in equation (1) has yielded a value of ΔE =12.3 keV, close to this reported value. It is needless to say that this extreme low energy difference between the two isomeric states is the reason for non-observance of isomeric transition between them.

***Table 3:*** *Physically admissible 2qp GM doublet bandheads in $^{154}$Tb from the coupling of 1qp proton (top row) and neutron orbital (first column) taken from $^{155}$Tb and $^{153}$Gd respectively. The entries in round brackets are summed ($E_p+E_n$) energies in keV.*

| $p_j \rightarrow$ <br> $n_i \downarrow$ | $p_0$  0 keV <br> $3/2^+$ [411↑] | $p_1$  227 keV <br> $5/2^-$ [532↑] | $p_2$  271 keV <br> $5/2^+$ [413↓] |
|---|---|---|---|
| $n_0$  0 keV <br> $3/2^-$ [521↑] | $3^-$  $0^-$ <br> (0) | $4^+$  $1^+$ <br> (227) | $1^-$  $4^-$ <br> (271) |
| $n_1$  58 keV <br> $3/2^+$ [651↑] | $3^+$  $0^+$ <br> (58) | $4^-$  $1^-$ <br> (285) | $1^+$  $4^+$ <br> (329) |
| $n_2$  95 keV <br> $1/2^+$ [660↑] | $2^+$  $1^+$ <br> (95) | $3^-$  $2^-$ <br> (322) | $2^+$  $3^+$ <br> (366) |
| $n_3$  110 keV <br> $5/2^-$ [523↓] | $1^-$  $4^-$ <br> (110) | $0^+$  $5^+$ <br> (337) | $5^-$  $0^-$ <br> (337) |
| $n_4$  129 keV <br> $3/2^-$ [532↓] | $0^-$  $3^-$ <br> (129) | $1^+$  $4^+$ <br> (356) | $4^-$  $1^-$ <br> (400) |
| $n_5$  171 keV <br> $11/2^-$ [505↑] | $7^-$  $4^-$ <br> (171) | $8^+$  $3^+$ <br> (398) | $3^-$  $8^-$ <br> (442) |

2. *The 22.7 h high spin isomer*

A third long-lived low-lying isomeric state reported in $^{154}$Tb is proposed to have $J^\pi=7^-$ through the decay studies. Activity studies of Lau & Hogan [41] suggest a small percentage of IT decay of this LLI to the low spin $J^\pi=3^-$ ($t_{1/2}$=9.4 h) isomer, while its energy remains unknown. As seen in Table 3, the coupling of the high-spin $\nu 11/2^-$ [505] orbital ($n_5$) with the $\pi 3/2[411]$ ($p_0$) results in $J^\pi=7^-$ state. The summed 1qp proton and neutron energies ($E_p+E_n$) suggest $E_x \geq 170$ keV for this $J^\pi = 7^-$ state. The 2qp-plus-phonon model calculations place the level at $E_x \approx$ 284 keV [18]. As seen in Table 3, all other 2qp bandheads arising from the coupling of ($p_0$, $n_i$) yield states with spin J≤4, to which gamma transitions from $J^\pi=7^-$ state will be highly hindered. All other 2qp bandheads from the coupling of ($p_1$,$n_i$) and ($p_2$,$n_i$) will have $E_x$ >200 keV owing to higher 1qp proton orbital energies. Hence, we confirm $J^\pi=7^-$ {$p_0$: 3/2[411] ⊗ $n_5$: 11/2[505]} as the third isomeric state, complying with the configuration proposed by Riedinger et. al. [38]. Listed in Table 3 are 18 pairs of physically admissible 2qp bandheads as determined from our TQRM analysis that are expected in the low-energy spectrum of $^{154}$Tb. It may be noted that many of these states have not been found experimentally to date.

In light of the above discussions, we propose the following assignments to the low-lying isomer triplet in $^{154}$Tb:

**21.5 h $^{154}$Tb $^{g.s}$:** $0^-$ {π3/2[411↑] ⊗ ν3/2[521↑]}; **$E_x$**=0 keV

**9.4 h $^{154}$Tb$^{m1}$:** $3^-$ {π3/2[411↑] ⊗ ν3/2[521↑]}; **$E_x$**≈12 keV

**22.7 h $^{154}$Tb$^{m2}$:** $7^-$ {π3/2[411↑] ⊗ ν11/2[505↑]}; **$E_x$** ≥ 170 keV

## C. Isomer triplet trend in light rare-earth odd-odd Tb isotopes

As analysed in the previous sections, both $^{154}$Tb and $^{156}$Tb are known to have low-lying long-lived isomer triplets with the same spin values: J = 0, 3 and 7. On comparison with the neighbouring odd-odd isotopes, it is observed that this trend exists in $^{158}$Tb as well, although the half-life of one of the isomers is relatively small. This trend is not seen in neighbouring higher mass or lower mass odd-odd Tb isotopes. The g.s. GM doublet is consistently formed by the coupling of $p_0$: 3/2[411↑] ⊗ $n_0$: 3/2[521↑] in $^{154,156,158,160}$Tb. While in $^{156,158,160}$Tb the g.s. is the spins anti-parallel triplet state, $K_T$ =3⁻ as expected, the spins parallel singlet partner $K_S$=0⁻ is experimentally reported to be the g.s. in $^{154}$Tb, in an apparent violation of the GM rule as discussed in the previous section. As an exception, in $^{156}$Tb, the $J^\pi$=0⁺{π3/2[411↑] ⊗ ν3/2[402↓]} forms the second isomeric state. As we move to $^{160}$Tb, the $K_S$=0⁻ no longer remains an isomer, as a $J^\pi$=1⁻ level exists below it facilitating an unhindered decay. A comparison of the experimentally observed low-lying 2qp bandheads in Tb isotopes is presented in Fig. 3.

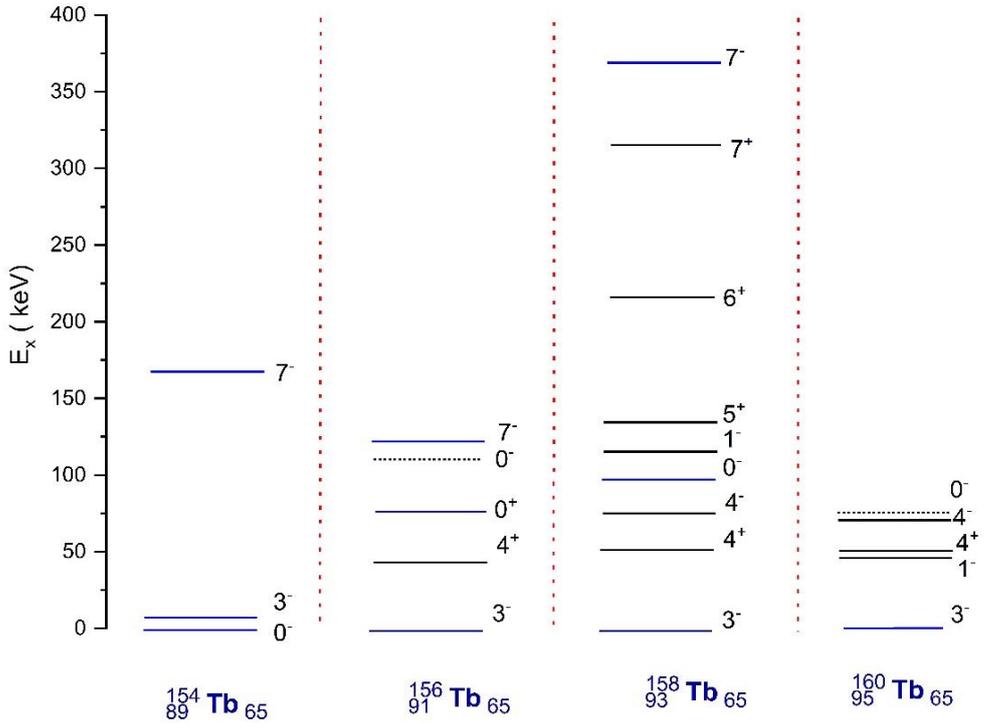

*Figure 3:* A comparison (not to scale) of the low-lying rotational levels of odd-odd Tb isotopes of interest in the present study. The J=0,3 and 7 states, highlighted in blue, form the isomer triplets in $^{154}$Tb, $^{156}$Tb and $^{158}$Tb.

The high spin isomer among the triplet observed in $^{154,156,158}$Tb is the $J^\pi$=7⁻{π3/2[411↑] ⊗ ν11/2[505↑]}. As seen in figure 1(a), the sharp drop in the energy of ν11/2[505↑] neutron orbital in the A=153-159 neighbouring Gd isotones results in the formation of this high spin state at low energies ($E_x$~300 keV). Further, all other physically admissible 2qp states within this energy range are generally of lower spins (J≤4), thereby making $J^\pi$=7⁻ an isomer. Even in $^{158}$Tb, the $J^\pi$ = 7⁻ state is a short-lived isomer, albeit lying close to $I^\pi K$= 7⁺4 rotational level, due to high K-hindrance factor. As the energy of the ν11/2[505↑] starts increasing sharply again beyond the A=159, the $J^\pi$=7⁻ level is no longer observed at energy below 400 keV, ending the isomer triplet trend.

## IV. Systematics of Isomer triplets in light-rare earth nuclei

With our understanding of their formation in the odd-odd Tb isotopes, a closer look at the characteristics of the isomer triplets in the doubly odd light rare-earth nuclei listed in Table 1 reveals certain interesting trends. Of the three isomers constituting the triplets, the g.s. and low spin LLI are found to have spins I = 0/1/2 and I=3/4/5 respectively or vice versa, while the third isomer is always of high spin with I=6/7/8/9. $^{166}$Lu is the only exception to this observed pattern. This feature is exclusive to nuclei in the transitional light rare-earth region. In the section that follows, we discuss the systematics of 1qp proton and neutron orbitals in the neighbouring nuclei thereby delineating the cause for the formation of low-lying LLIs in these nuclei.

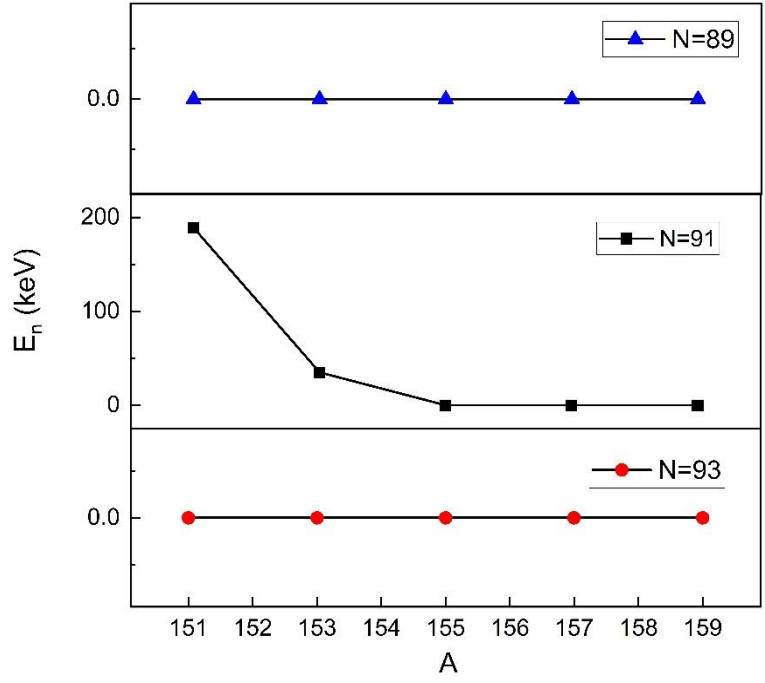

*Figure 4: A comparison of energy systematics of the single particle neutron orbital ν3/2[521] in N=89,91 and 93 isotones in the 150<A<160 mass region.*

### A. Ground state and the low-spin LLI

In a trend similar to the Tb isotopes, we find the ν3/2[521↑] single particle neutron orbital to be foremost in the formation of the lowest two LLIs in the other isomer triplets listed in Table 1. The involvement of this n-orbital in the isomer formation is understood as it consistently remains as g.s in all N=93 isotones in the entire mass region of A=150-160 [43] and in N=89 and 91 isotones in the mass region 153 ≤ A ≤ 160.

#### 1. $^{152}$Pm

The latest NDS [20] lists the g.s. orbital configuration of $^{152}$Pm to be $J^\pi$=1$^+$\{π5/2[532↑] ⊗ ν3/2[532↓]\}. This $J^\pi$ and configuration of the g.s. was proposed through the observed unhindered β-decay from g.s. of $^{152}$Nd [44]. The ν3/2[532↓] orbital is expected to form the g.s. and low-spin LLI in $^{152}$Pm as it is observed to be below the ν3/2[521↑] for the N= 91 isotones in A<153 region. The two lowest lying 1qp proton orbitals in the neighbouring isotopes of $^{152}$Pm are π5/2[532↑] and π5/2[413↓]. The former is reported to be the g.s. proton orbital throughout the A=151-159 Pm odd-mass isotopes and thereby couples with previously discussed ν3/2[532↓] n-orbital to give $J^\pi$=1$^+$ as the g.s. of $^{152}$Pm. The π5/2[413↓] orbital on the other hand, is a part of the g.s. parity-doublet in $^{151}$Pm followed by a sharp increase in its energy in the heavier isotopes. The energy trend of these two proton orbitals is depicted in Fig 5, as taken from the NDS [20]. The low-spin isomer of $^{152}$Pm has been assigned $J^\pi$=4$^-$(7.52 min), its spin assigned unambiguously from β-decay studies [45], albeit without its configuration. The decay study also reports the isomer's β-decay to the g.s. from which its energy is assigned as $E_x$≈150 keV. Through our TQRM analysis, we propose the following configuration for this isomer consistent with the 1qp orbital systematics in the neighbouring nuclei:

**7.5 min $^{152}$Pm$^{m1}$:** 4$^-$\{π5/2[413↓] ⊗ ν3/2[532↓]\}

It may be noted that the spins-antiparallel singlet partner of the g.s GM doublet, $K_s$=4$^+$\{π5/2[532↑] ⊗ ν3/2[532↓]\} is expected to lie 60 keV above it.

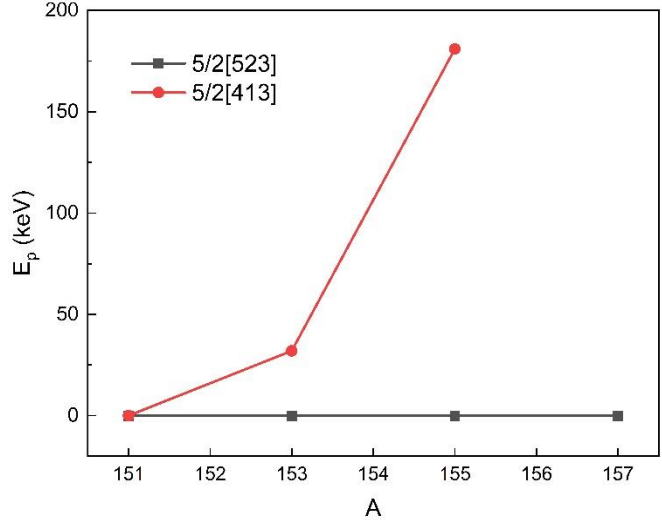

*Figure 5*: *Energy systematics of the low energy single particle proton orbitals 5/2[532] and 5/2[413] in the A=151-159 Pm isotopes*

### 2. $^{152}$Eu

The g.s. of $^{152}$Eu is confirmed by Lanier et. al. [46] to be $J^{\pi}= 3^-\{\pi5/2[413\downarrow] \otimes \nu11/2[505\uparrow]\}$ through experimental studies of its g.s. rotational structure. The study reported the g.s. to have a high deformation which supports an unusually stable rotational band up to $J^{\pi}=7^-$ level. The 9.3 h low-spin isomer was assigned spin $J^{\pi}=0^-$ from its β-decay to the deformed low spin levels of $^{152}$Gd and $^{152}$Sm [47]. The isomer is placed at $E_x$=45 keV and is reported to undergo 100% independent β-decay. However, its orbital configuration is so far undetermined. Lying close to the transition region, the 1qp orbital systematics of the Eu isotopes seem to vary from the Nilsson model ordering as seen in the case of its highly deformed g.s.. Also, the lack of available data on the low-lying $^{151}$Eu orbitals makes the choice of 1qp proton orbitals for the formation of 2qp states uncertain. This isotopic neighbour is known to exhibit co-existence of spherical and deformed shapes. The adopted g.s. spin value of $^{151}$Eu is the $d_{5/2}$ orbital 5/2[402], which is not seen as the g.s. of its following higher mass isotopes. This and the reported 2qp g.s. configuration of $^{152}$Eu involving the usually high energy 11/2[505] orbital proves that close to the transitional region, drastic changes in shape and structural properties can be anticipated. The predicament of some states accounted for by the shell model and some by the Nilsson model for odd-A nuclei is discussed in detail by Jain et. al. [48].

According to the systematics available for higher mass isotopes (A≥ 153), the contenders for the lowest lying p orbitals are π5/2[413], π5/2[532] and π3/2[411]. Our systematics study of N=89 isotones has already established the ν3/2[521] as the g.s. 1qp n orbital. The other identified low-lying states are ν3/2[532] and ν5/2[523], both of which form the g.s. in the isotonic neighbour $^{151}$Sm, but rapidly rise with increase in mass number. From the available systematics of 1qp n and p orbitals, we can see that only the coupling of π3/2[411] orbital with either ν3/2[521] or ν3/2[532] can give rise to a $J^{\pi}$=0$^-$ state at energies as low as $E_x$~50 keV. The coupling of π3/2[411] ⊗ ν3/2[521] forms $K_S$=0$^-$, suggesting a higher energy placement while in π3/2[411] ⊗ ν3/2[532] the 2qp state formed is $K_T$=0$^-$, a low-lying triplet state. However, as discussed before, the energy systematics of these 1qp orbitals close to the transitional region are subject to abrupt changes. Hence, the orbital configuration of the low spin LLI $J^{\pi}$=0$^-$ could be either of the above two.

### 3. $^{156,158,160}$Ho

The trend of the ν3/2[521↑] neutron orbital confirms its involvement in the formation of g.s. and low spin isomer in $^{156,158,160}$Ho. Through collinear spectroscopy investigations [43], the g.s. of $^{156}$Ho was found to have $J^{\pi}$=4$^-$spin. Its configuration was proposed to be π5/2[402↑] ⊗ ν3/2[521↑] based on the 1qp orbital systematics and the GM rule [20]. Its singlet partner $J^{\pi}$=1$^-$ {π5/2[402↑] ⊗ ν3/2[521↑]} form the low spin isomer ($t_{1/2}$=9.5 s) at energy $E_x$=57.4 keV. This state is believed to undergo a 100% IT decay to the g.s. as no β/e.c. decay has been observed.

Beyond $^{156}$Ho, there is crossover of the two Nilsson orbitals π5/2[402↑] and π7/2[523], thereby resulting in a change in the g.s. in higher mass Ho isotopes, a trend that has been established till A=170. Thus, the g.s. of $^{158,160}$Ho have the same spin and orbital configuration viz. $J^π$=5$^+${π7/2[523↑] ⊗ ν3/2[521↑]}. The low-spin isomer of $^{158}$Ho ($t_{1/2}$=28 min) is listed in the NDS [20] to be a $J^π$=2$^-$ state at energy Ex=67keV. Its orbital configuration is considered to be π7/2[404↓] ⊗ ν3/2[521↑] based on TQRM calculations by Sood et. al. [49]. Atomic spectroscopy and observed E3 IT decay to the g.s. confirm the low spin isomer ($t_{1/2}$=5.02 h) of $^{160}$Ho to also be $J^π$=2$^-$ at energy $E_x$= 60 keV. Our observation of 1qp proton systematics suggest that the orbital configuration of this state will also be similar to that of $^{158}$Ho. We hence propose the following assignment:

**5.02 h $^{160}$Ho$^{m1}$:** 2$^-$ {π7/2[404↓] ⊗ ν3/2[521↑]}

## 4. $^{162,166}$Lu

The g.s. spin-parity of $^{162}$Lu was determined through laser collinear spectroscopy [50] as $J^π$=1$^-$ with an expected configuration π1/2[411↓] ⊗ ν3/2[521↑]. However, the spin-parity and energy of its low spin LLI ($t_{1/2}$=1.5 min) is unknown. The trend of g.s. and low spin LLI is completely different in $^{166}$Lu compared to the other nuclei of interest in the present study. Hence, the low-lying isomers of these Lu isotopes is being pursued as a separate study to understand the systematics.

## 5. ΔI=3 isomer enigma

A striking observation brought out by the above discussions is that the spin difference between the g.s and the low-spin isomer in all the triplet nuclei is ΔI=3. In these cases, an M3/E3 transition can be expected to de-excite the nucleus to the g.s. Their lesser transition probability compared to other lower order multipolarities could conceivably give rise to the spin isomerism of the LLI. Experimentally, M3/ E3 transitions between the low-spin LLI and g.s. have been observed only in $^{156}$Tb, $^{156}$Ho, $^{158}$Ho, $^{160}$Ho and $^{166}$Lu. Indirect experimental evidence of isomeric transition in $^{154}$Tb has been reported, as mentioned before. Here, the proposed low energy difference between the low-spin LLI and the g.s makes the $J^π$= 3$^-$ an extremely low energy (ELE) isomer. The hindrance in the E3 decay in this case is also caused by the dependence of the γ-decay probability on the transition energy [8]. In the other three nuclei, viz. $^{152}$Pm, $^{152}$Eu and $^{162}$Lu, the low-spin isomer is reported to undergo 100% β-decay. Previous studies have attributed this absence of isomeric transition in $^{152}$Eu to the relatively higher deformation of g.s. [18,46], while no data is available regarding the low spin LLI in the other two nuclei. The lacuna can be attributed to the lack of dedicated experimental studies of their low energy spectrum since their discovery. This warrants specific experimental studies and theoretical studies with a framework including their transitional matrix element to explore any other possible structural effects causing the isomerism.

### B. High-spin isomer

The NDS [20] listed spin-parities and orbital configurations of the high-spin LLI in the isomer triplets of our interest except the Lu isotopes are presented in Table 4. As clearly observed, all these states are formed by the coupling of the single particle n orbital ν11/2[505] with relevant p orbitals. The isomer trend in $^{162,166}$Lu do not involve the ν11/2[505] orbital and hence are subject of independent study, as mentioned before. This high spin neutron state is generally observed at higher energies in other mass regions. But in the transitional light rare-earth region, its energy steeply drops, leading to its coupling with low-lying proton states and eventually forming high spin 2qp states at lower energies. This low energy trend of ν11/2[505] can be understood through the Nilsson model energies [1,2]. In the N=89,91 and 93 isotones, this component of $1h_{11/2}$ lies closer to the Fermi surface in the transitional light rare-earth region. Hence, neutron excitation to this high spin intruder orbital is possible at low excitation energies. In Fig. 5, we have mapped the systematics of the ν11/2[505] orbital in the N= 89,91 and 93 isotones in the light rare-earth region to highlight its significance in the formation of the high spin LLIs. As seen therein, in the N=89 and 91 isotones, this n-orbital is found below 200 keV in the A=153-157 mass region. Consequently, its coupling with the corresponding low-lying p-orbitals results in the high-spin isomeric state of $^{152}$Eu, $^{154}$Tb (N=89 isotones) and $^{156}$Tb,$^{158}$Ho (N=91 isotones). Similarly, in the N=93 isotones, a drop in the energy of ν11/2[505] occurs around A= 159, thereby giving rise to the high spin isomer in $^{160}$Ho. We may also note that this neutron orbital is also involved in the high spin isomeric state of $^{154}$Eu ($E_x$= 145 keV; $J^π$ =8$^-$) and $^{158}$Tb ($E_x$ =

388keV ; $J^\pi =7^-$), though these nuclei do not exhibit isomer triplets. In all these, the ν11/2[505] state couples with the g.s. 1qp proton orbitals in respective nuclei leading to low-lying LLIs.

*Table 4: List of high spin I=7/8/9 LLIs and their respective orbital configurations in light rare-earth nuclei with isomer triplets as adopted and listed in the NDS [20]. Configuration of the $^{158}$Ho LLI was proposed in a previous TQRM study [49].*

| Nucleus | High spin $J^\pi$ | Orbital Configuration |
|---|---|---|
| $^{152}_{61}$Pm$_{91}$ | (8) | --- |
| $^{152}_{63}$Eu$_{89}$ | $8^-$ | π5/2[413] ⊗ ν 11/2[505] |
| $^{154}_{65}$Tb$_{89}$ | $7^-$ | π3/2[411] ⊗ ν 11/2[505] |
| $^{156}_{65}$Tb$_{91}$ | $7^-$ | π3/2[411] ⊗ ν 11/2[505] |
| $^{156}_{67}$Ho$_{89}$ | $9^+$ | π7/2[523] ⊗ ν 11/2[505] |
| $^{158}_{67}$Ho$_{91}$ | $(9^+)$ | π7/2[523] ⊗ ν 11/2[505] |
| $^{160}_{67}$Ho$_{93}$ | $(9^+)$ | π7/2[523] ⊗ ν 11/2[505] |

In $^{152}$Pm, the high spin member of the isomer triplet is tentatively assigned a spin of J=8. This isomeric state was first reported by Daniel et al., in 1971 in their beta decay studies of $^{152}$Nd and $^{152}$Pm [45]. They had suggested a higher spin for this isomer as it primarily beta decays to high spin states in $^{152}$Sm. In the current data sheets for $^{152}$Sm, the NDS evaluators have commented that the probable spin for this isomer is $J^\pi=8^-$based on the observation of a 229 keV γ-ray from the $8^-$ level in $^{152}$Sm, fed by the β-decay from this 13.8 min isomer [20]. Based on the single particle proton and neutron energy systematics in the neighbouring odd-mass nuclei and the reported spin values, we propose the possible orbital configurations for this high spin LLIs as:

**13.8 min $^{152}$Pm$^{m2}$**: $J^\pi=8^-$ {π5/2[413] ⊗ ν11/2[505]}.

The high spin isomers of $^{152}$Eu, $^{156}$Tb and $^{160}$Ho are observed to undergo 100% IT decay, in the first two nuclei through an E3 isomeric transition and in the third through an M2 transition. IT decays in other four nuclei, $^{152}$Pm, $^{154}$Tb, $^{156}$Ho and $^{158}$Ho have not been directly observed so far, as the expected transitions are of higher multipolarity and are hence less probable. They are however expected to have a small percentage of IT decay to the lower energy states as suggested by theoretical calculations and β-decay studies [20]. Based on the defined classifications of isomers, the high spin LLI of the triplet is a combination of predominantly spin isomerism with a contribution of K-hindrance in some nuclei. The I=7/8/9 high spin LLI has no level with spin difference ΔI<4 below it for an unhindered γ-decay. The high spin isomers of $^{152}$Eu, $^{156}$Tb and $^{160}$Ho are purely spin isomers as their K-hindrance factor ν=1. In the other nuclei of interest, the contribution of K-hindrance alongside the spin isomerism is unknown due to ambiguities in their respective low-lying level structures.

In the updated Atlas of Isomers, Jain et. al. [13] report that the highest number of high spin isomers are observed to have spins $J^\pi= 11/2^-$and $13/2^+$ in odd-mass nuclei and $J^\pi= 7^-, 8^-$and $8^+$ in case of even-mass nuclei. These studies have brought forth the role played by shell model single particle orbital systematics in the formation of spin isomers in the vicinity of magic nuclei and the proton drip line caused by high spin intruder $g_{9/2}$, $h_{11/2}$ and $i_{13/2}$ orbitals. The observation of spin isomer formations in the odd-odd nuclei of the light rare-earth region reported and analysed in our study is attributed to the single particle neutron orbital systematics at low energies. Going by the Nilsson model, the energy systematics of the ν13/2[606] single particle neutron orbital can be mapped, similar to our study of the ν11/2[505] orbital, to identify the mass regions where it can form high spin isomers in odd-odd nuclei. Similar investigation of high spin single particle proton orbital trends can be performed to predict the possible occurrences of high spin isomers at low energies. Trends of such high spin single particle orbitals which potentially could form high spin isomers can be traced across mass regions in case of even-even and odd-mass nuclei as well.

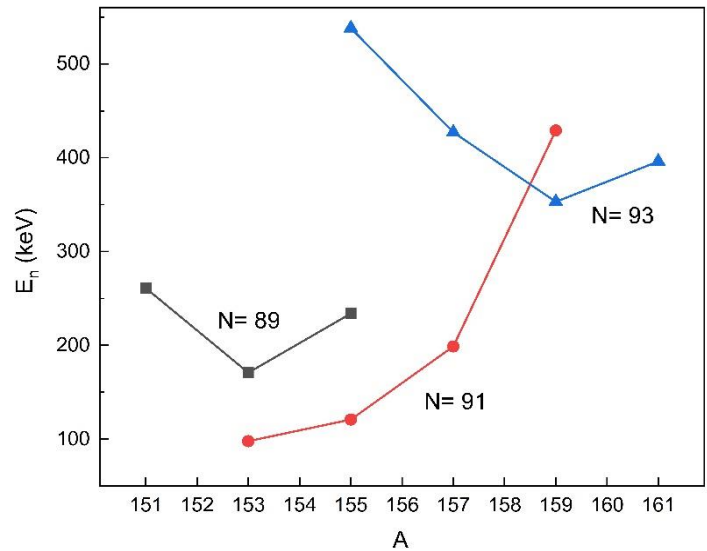

*Figure 6: Energy systematics of the ν11/2[505] single particle neutron orbital in the light rare-earth mass region A=150-160 in N=89,91 and 93 isotones*

## V. Summary

An extensive systematic study of long-lived isomers (LLIs) across the nuclear landscape has brought out an interesting trend of low-lying isomer triplets in the N=89,91 and 93 isotones in the light rare-earth region. The Two Quasiparticle Rotor Model has been used to investigate the level structures of $^{154}$Tb and $^{156}$Tb with a particular focus on their low-lying isomer triplets. The physically admissible 2qp bandheads of these odd-odd nuclei were determined by studying the systematics of proton and neutron orbitals in neighbouring Tb isotopes and N=89 and N=91 isotones in the 150 <A <160 mass region. Our calculations have proposed 15 and 18 admissible low-lying 2qp states in $^{156}$Tb and $^{154}$Tb respectively. The J$^\pi$ and configurations of the isomers forming the triplet in both the nuclei were determined and their energies justifying their isomeric nature were proposed with reference to the adopted experimental data. There is a need for dedicated experiments aimed at studying the low-lying level structure of both these isotopes for which the levels obtained from our calculations can serve as location guides.

Our extended study on the systematics of isomer triplets in neighbouring odd-odd Tb isotopes as well as other rare-earth nuclei of interest in the A=152-160 mass region revealed a unique recurring pattern of spins and orbital configurations. A detailed review of low-lying 1qp proton and neutron orbitals in the rare-earth region suggests a drastic change in their energy systematics close to the transitional region (A~150). The 1qp neutron orbital trend of the lowest lying ν3/2[521], its coupling with the respective low-lying proton orbitals and the presence of the high spin ν11/2[505] orbital at low energies are found to be causing this fascinating trend of isomer triplets. We have suggested the orbital configurations and energies for some of the low-lying LLIs from our analysis. The absence of a ΔI=3 isomeric transition from the low spin isomer to the g.s. remains an unsolved problem as pointed out previously in the Atlas of Isomers [13] and Jain et. al. [18]. This puzzling phenomenon is found to be in some other odd-odd rare earth nuclei as well and warrants dedicated theoretical and experimental studies for comprehensive understanding. Our study of this unique trend of isomer triplets and their energy systematics is an attempt to highlight the plethora of structural anomalies found close to the transitional region. Further studies on these nuclei, especially farther from the region of stability, can unravel and elucidate their exotic properties, which in turn will enrich the overall understanding of nuclear structure.

**Acknowledgement:** This work is dedicated to our Revered Founder Chancellor Bhagawan Sri Sathya Sai Baba. The authors deeply acknowledge the discussions and valuable inputs from Prof. P. C Sood.

# VI. References

[1] A.Bohr & B.R. Mottelson, Nuclear Structure Volume II: Deformations, World Scientific Publishing Co. Pte. Ltd., Singapore, 1998.

[2] R. F. Casten, Nuclear Structure from a Simple Perspective, Oxford Science Publications, New York, 1999.

[3] D. Bonatsos et. al., Signatures for shape coexistence and shape/phase transitions in even-even nuclei, Journal of Physics G: Nuclear and Particle Physics 50, (2023).

[4] A. Mukherjee et al., Evidence of transverse wobbling motion in $^{151}$Eu, Phys. Rev. C 107, (2023).

[5] D. J. Hartley et al., Wobbling mode in $^{167}$Ta, Phys. Rev. C 80, (2009).

[6] S. Nandi et al., First Observation of Multiple Transverse Wobbling Bands of Different Kinds in $^{183}$Au, Phys. Rev. Lett. 125, (2020).

[7] G. B. Hagemann et al., Evidence for the wobbling mode in nuclei, Phys. Rev. Lett. 86, 5866 (2001).

[8] A. K. Jain, B. Maheshwari, and A. Goel, Nuclear Isomers A Primer, Springer Nature, Switzerland 2021.

[9] P. M. Walker et. al., Nuclear Isomers, Eur. Phys. J. Spec. Top. (2024) 233:889–892.

[10] A. Bohr, Rotational motion in nuclei, Rev Mod Phys 48, (1976).

[11] A. Bohr and B.R Mottelson., Rotational States in even-even nuclei, Phys. Rev. 90, 717 (1953).

[12] P. Walker & Z. Podolyák, 100 years of nuclear isomers - Then and now, Phys. Scr. 95 (2020) 044004.

[13] S. Garg, et. al., Atlas of Nuclear Isomers-Second Edition, At. Data Nucl. Data Tables 150, 101546 (2023).

[14] G. D. Dracoulis, et. al., Review of Metastable States in Heavy Nuclei, Rep. Prog. Phys. 79 (2016) 076301.

[15] B. Maheshwari & A. K. Jain, Nuclear Isomers at the Extremes of Their Properties, Eur. Phys. J. Spec. Top. (2024) 233:1101–1111

[16] P. C. Sood & R. K. Sheline, Long-lived isomers in medium-heavy and heavy deformed nuclei, Nucl. Inst. and Meth. in Phys. Res. B24/25 (1987) 473-476.

[17] R. Orford et al., Spin-trap isomers in deformed, odd-odd nuclei in the light rare-earth region near N=98, Phys. Rev. C 102, 011303 (2020).

[18] A. K. Jain et. al., Nuclear Structure in Odd-Odd Nuclei, 144≤A≤194, Rev. Mod. Phys., 70, (1998).

[19] D. M. Headly et. al., Intrinsic structures and associated rotational bands in medium-heavy deformed odd-odd nuclei, At. Data Nucl. Data Tables 69, 239–348 (1998).

[20] Evaluated Nuclear Structure Data File (ENSDF) and XUNDL (Current Version), continuously updated data files (NNDC, Brookhaven, NY).

[21] P. C. Sood and R. Gowrishankar, Configuration assignments to isomers in the neutron-rich $^{186}$Ta (Z=73) nucleus, Phys. Rev. C 90, (2014).

[22] P.C. Sood et. al., Intrinsic and Rotational Level Structures in Odd-Odd Actinides, At. Data Nucl. Data Tables 58 (1994).

[23] C.J. Gallagher & S.A. Moszcowski, Coupling of Angular Momenta in Odd-Odd Nuclei, Phys. Rev. 111, (1958).

[24] P. C. Sood et. al., Level structures of the transfermium odd-odd nucleus $^{252}$Md, Phys. Rev. C 103, (2021).

[25] P. C. Sood and R. Gowrishankar, Low-lying level structures in the transuranic n -rich Z=93 nuclei $^{243}$Np and $^{244}$Np, Phys Rev C 106, (2022).

[26] R. Gowrishankar and P. C. Sood, Level structures in odd-odd deformed nucleus $^{184}$Ta, Eur. Phys. Jour. A 52, (2016).

[27] P. C. Sood et. al., Level structures in the odd-odd nucleus $^{154}$Pm, J. of Phys. G: Nucl. and Part. Phys. 39, (2012).

[28] P.C. Sood et. al., Level structures in $^{156}$Pm from $^{156}$Nd β$^−$ decay, Eur. Phys. J. A 48, (2012).

[29] P.C. Sood et. al., Level structures in $^{240}$Np, Phys. Rev. C 89, 034308 (2014)

[30] P.C. Sood and R. Gowrishankar, Characterization of long-lived isomers in the odd-odd heavy actinide $^{254}$Md, Phys. Rev. C 95, 024317 (2017)

[31] K. E. Ådelroth et. al., Nuclear Spins of Neutron-Deficient Terbium Isotopes, Phys. Scr. 2, (1970).

[32] J. W. Mihelich et. al., Nuclear spectroscopy of neutron-deficient rare earths (Tb through Hf), Phys. Rev. 108, 989 (1957).

[33] J. W. Mihelich & B. Harmatz, Some new isomeric transitions in rare earth nuclei, Phys. Rev. 106, 1232 (1957).

[34] T. Toriyama et. al., Existence of a new isomer of T½=24.4 hr in$^{156}$Tb, J. Physical Soc. Japan 29, 9 (1970).

[35] P. C. Sood et. al., Level Structures in the N=91 Odd-Odd Nucleus, Proceedings of the DAE-BRNS Symp. on Nucl. Phys. 60 (2015)

[36] R. Bengtsson et. al., High-spin states in the odd-odd $^{154}$Tb and $^{156}$Tb nuclei and the systematic for the [i$_{13/2}$]$_n$[h$_{11\ 2}$]$_p$ bands, Nucl. Phys. A 389, (1982).

[37] D. C. Sousa et. al., Decay of the three isomers of $^{154}$Tb, Nucl. Phys. A 238, (1975).

[38] L. L. Riedinger, et. al., Decay of a new isomer in $^{154}$Tb to high-spin levels in $^{154}$Gd, Phys. Rev. C 4, 1352 (1971).

[39] J. Ferencei et. al., Nuclear orientation of $^{152,154}$Tb in Gadolinium, Czech. J. Phys. B 31 11981.

[40] Gy. Gyürky et. al., Precise half-life measurement of the 10 h isomer in $^{154}$Tb, Nucl. Phys. A 828 (2009).

[41] J. C. F. Lau & J. J. Hogan, Investigation of branching ratios between isomeric states of $^{154}$Tb, Phys. Rev. C 8, 715 (1973).

[42] B. Harmatz et. al., Nuclear Levels in a Number of Even-Even Rare Earths (150 <A <184), Phys. Rev. 123, (1961).

[43] G. S. Simpson et al., Near-yrast structure of N=93 neutron-rich lanthanide nuclei, Phys. Rev. C 81, (2010).

[44] M. Shibata et. al., Decay Scheme of Mass-Separated $^{152}$Nd, Appl. Radiat. Isot. 44, (1993).

[45] W. R. Daniels & D. C. Hoffman, Decay of $^{152}$Nd and the isomers of $^{152}$Pm, Phys. Rev. C 4, 919 (1971).

[46] R. G. Lanier et. al., On the stability of a ground-state rotational structure for the transitional nucleus $^{152}$Eu, Phys. Lett. 78B (1978).

[47] O. Nathan & M. A. Waggoner, Of decay Eu$^{152}$ and Eu$^{152m}$, Nuclear Physics 2 (1966)

[48] A. K. Jain et. al., Intrinsic States of Deformed Odd-A Nuclei in the Mass Regions (151 ∼ A ∼ 193) and ( A ∼ 221), Rev. Mod. Phys. 62, (1990).

[49] P. C. Sood et. al., Characterization of Isomers in $^{158}$Ho, Phys. Rev. C 33, (1986).

[50] U. Georg et. al., Laser Spectroscopy Investigation of the Nuclear Moments and Radii of Lutetium Isotopes, Eur. Phys. J. A 3, (1998).